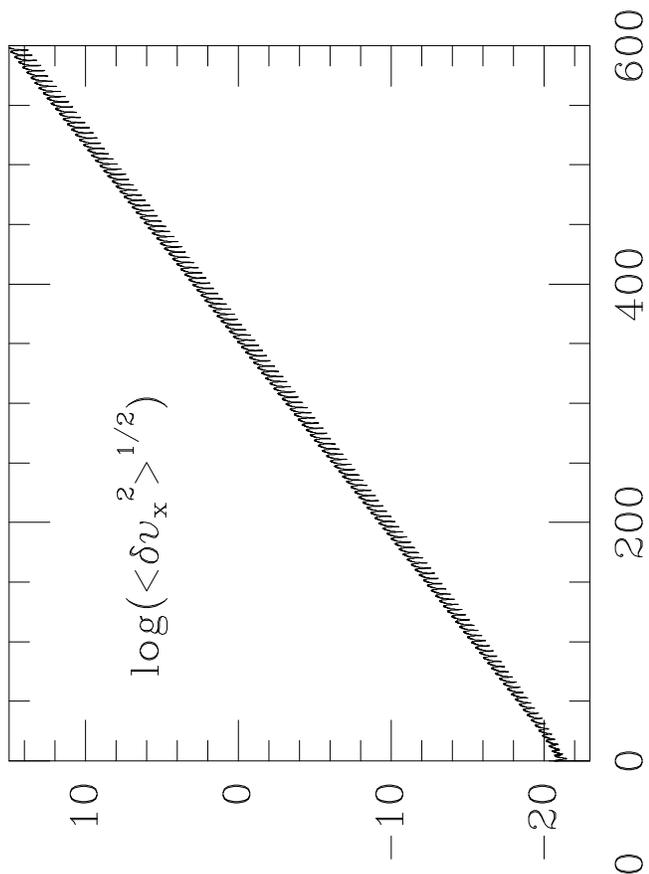

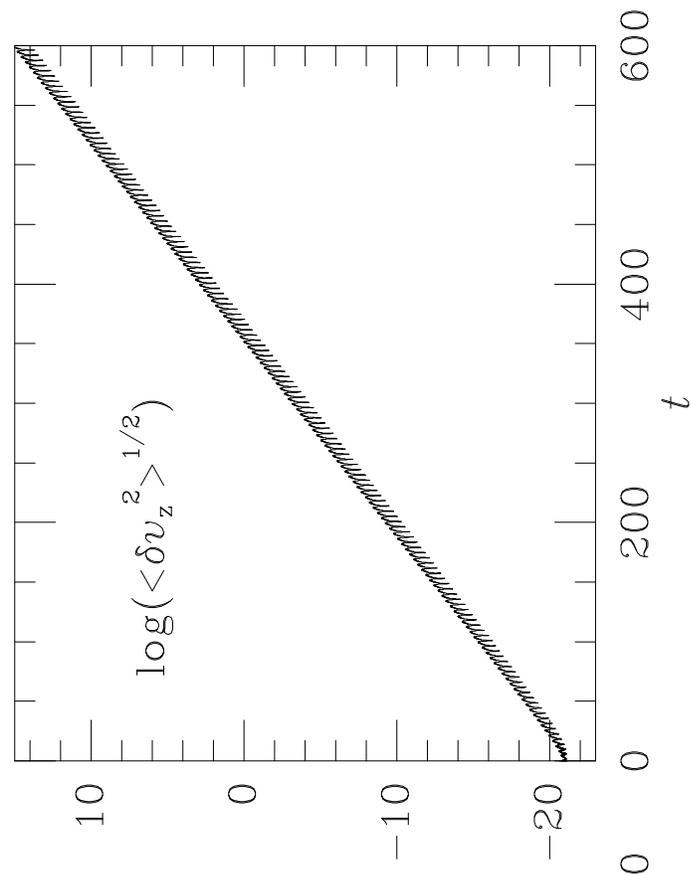

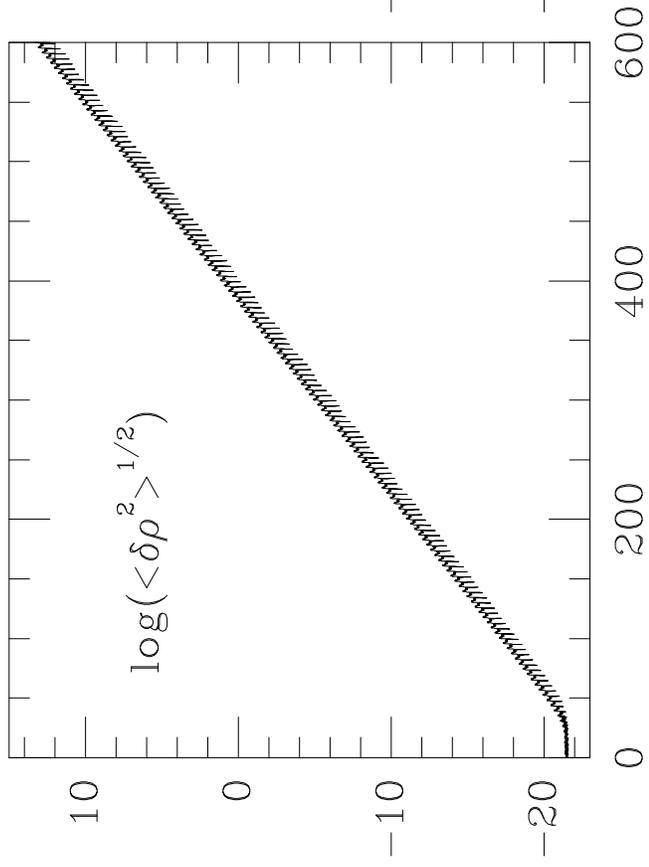

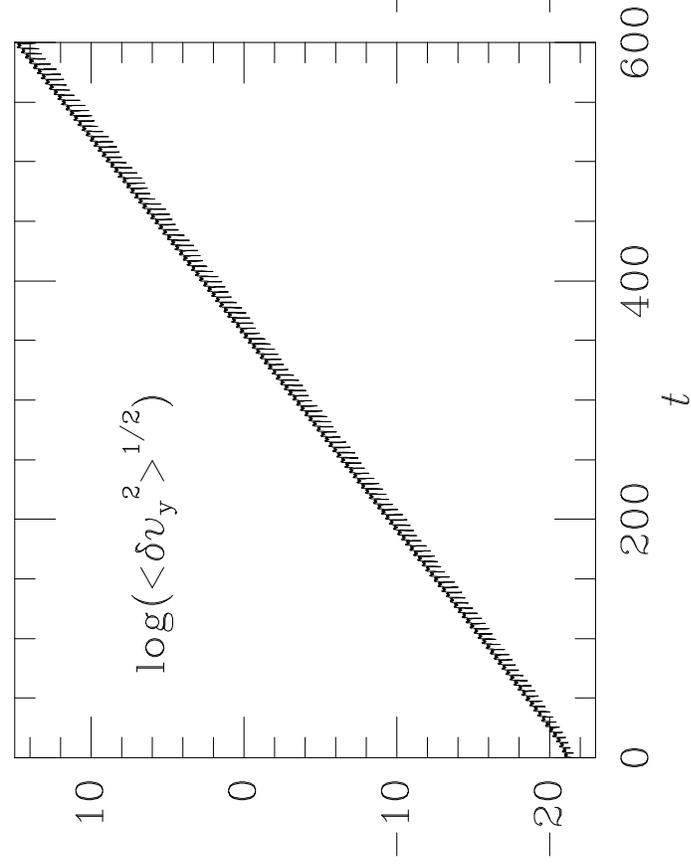

# Global Aspects of Elliptical Instability
# in Tidally Distorted Accretion Disks[1]

Dongsu Ryu

Department of Astronomy and Space Science, Chungnam National University, Korea

Jeremy Goodman

Princeton University Observatory, Peyton Hall, Princeton, NJ 08544

Ethan T. Vishniac

Department of Astronomy, University of Texas, Austin, TX 78712

## ABSTRACT

Tidally distorted accretion disks in binary star systems are subject to a local hydrodynamic instability which excites $m = 1$ internal waves. This instability is three dimensional and approximately incompressible. We study the global aspects of this local instability using equations derived under the shearing sheet approximation, where the effects of the azimuthal variation along distorted orbital trajectories are included in source terms which oscillate with local orbital phase. Linear analyses show that the excitation of the instability is essentially local, i.e. insensitive to radial boundary conditions. The region of rapid growth feeds waves into the region of slow or negligible growth, allowing the instability to become global. The global growth rate depends the maximum local growth rate, the size of the rapid growth region, and the local group velocity. We present an empirical expression for the global growth rate. We note that the local nature of the instability allows the excitation of waves with $m \neq 1$ when the local growth rate is large.

*Subject headings:* accretion, accretion disks − hydrodynamics − instabilities

---

## 1. Introduction

The evolution of accretion disks is governed by the transport of angular momentum outward, which allows mass to flow inward (see Pringle 1981 and Frank, King, & Raine 1992 for review and further references). In the standard $\alpha$-disk model (Shakura & Sunyaev 1973; Lynden-Bell & Pringle 1974), it is assumed that angular momentum transport is due to a local parameterized viscosity of the form

$$\nu = \frac{\alpha H^2}{\Omega},\tag{1}$$

where $H$ and $\Omega$ are the disk thickness and the orbital frequency respectively and $\alpha$ is a constant of order unity. Since microscopic molecular and radiative viscosities are normally very small, turbulence is often invoked as the source of the local viscosity. Consequently, local instabilities capable of inducing turbulence play a critical role in the theory of accretion disks.

Recently, Goodman (1993, henceforth G93) identified a hydrodynamic instability caused by tidal distortions of accretion disks. This instability follows from earlier work showing that laminar flows with high symmetry following rectilinear and circular paths may be stable against the growth of linear perturbations, but small departure from symmetry may lead to instabilities (Bayly, Orszag, & Herbert 1988 and references therein). The original work on this idea was done in the context of inviscid incompressible elliptical vortices, so the instability is called the *elliptical instability* (Gledzer et al. 1974; Gledzer et al. 1975; Pierrehumbert 1986; Bayly 1986; Craik & Criminale 1986; for further references, see Lubow, Pringle, & Kerswell (1993)). Since the angular momentum of the excited waves has the opposite sign of the specific angular momentum of the gas in the disk, and of the stars in the binary system, this process leads to the transfer of angular momentum from the disk to the binary system. In that role it is important in limiting the outer radius of the accretion disk and in providing a sink for the angular momentum transported outward from small radii.

At smaller radii within the disk the direct excitation of waves by the elliptical instability becomes weaker, and eventually completely ineffective. Nevertheless it may still play an important role in angular momentum transport in the inner parts of the disk. In unmagnetized disks the inward traveling waves excited at large radii can redistribute angular momentum, driving the evolution of the disk (Vishniac & Diamond 1989). In magnetized disks the local transport of angular momentum is probably mediated by a general magnetohydrodynamic instability first discovered by Velikhov (1959) (see also Chandrasekhar 1961) whose importance in the context of accretion disks was pointed out by Balbus & Hawley (1991). This process requires only that the disk be magnetized, with a magnetic pressure not greater than the ambient gas pressure, and with an angular velocity that decreases away from rotation axis. Balbus and Hawley proposed that this process is capable of driving a dynamo with a growth rate comparable to $\Omega$, a conjecture supported by recent numerical calculations (Brandenburg et al. 1995; Hawley, Gammie, & Balbus 1995). One of us (ETV) has argued elsewhere (e.g. Vishniac & Zhang (1995)) that this conclusion is inconsistent with the evidence for variations of $\alpha$ with temperature and radius found in phenomenological models of dwarf novae and X-ray transients. In the absence of a local dynamo, the waves excited by the elliptical instability are also capable of driving a dynamo (Vishniac, Jin, & Diamond 1990, Vishniac & Diamond 1992).

The linear analysis of the elliptical instability in accretion disks was originally performed by G93. In binary systems, disks are tidally distorted in a predominantly $m = 2$ pattern. These tidal distortions can destabilize $m = 1$ inertial waves, whose restoring force is due to the gradient of the specific angular momentum in the disk. In real accretion disks the relevant modes are strongly affected by the vertical entropy gradient as well and are frequently referred to as internal waves, by analogy with underwater ocean waves. G93 found that the local linear growth rate at small radii ($r \ll r_{\rm b}$) is

$$s_{\rm loc} \approx \frac{15}{4} \frac{M_{\rm s}}{M_{\rm p} + M_{\rm s}} \frac{\Omega_{\rm b}^2}{\Omega}, \tag{2}$$

where $M_{\rm p}$ and $M_{\rm s}$ are the mass of primary and secondary stars, $\Omega_{\rm b} = [G(M_{\rm p} + M_{\rm s})/r_{\rm b}^3]^{1/2}$ is the orbital frequency, and $r_{\rm b}$ is the binary semi-major axis. This formula can underestimate the growth rate when $r/r_{\rm b}$ is not small.

The nonlinear behavior of the instability was studied numerically by Ryu & Goodman (1994, henceforth RG) using a quasi-axisymmetric approximation in the meridional ($rz$) plane. For perturbation wavelengths small compared with the disk thickness, unstable modes were shown to remain nearly incompressible even when nonlinear. The modes saturated chaotically at perturbation velocity

$$\delta v \approx (11 \pm 2) \frac{s_{\rm loc}}{k}, \tag{3}$$

where $2\pi/k$ is the largest unstable wavelength in the simulations. The two-dimensional power spectrum of velocities after saturation was roughly isotropic and extended over a broad range of scales in an approximately power-law fashion.

The significance of this instability hinges on its persistence in the face of realistic damping mechanisms. This is dependent both on the local growth rate, and on the sensitivity of the growth rate to boundary conditions. Lubow, Pringle, & Kerswell (1993) argued that the elliptical instability is likely to prove insignificant on both grounds. They calculated the growth rate for large vertical wavelength modes, including the effects of vertical structure and stratification, and derived growth rates a factor of ten smaller than the asymptotic formula given in equation (2). They also estimated the effects of energy losses at disk boundaries, and concluded that these could result in the suppression of the instability. Subsequent work by Vishniac & Zhang (1995) has shown that the inclusion of realistic vertical structure does not substantially reduce the growth rates, and that the reduction found by Lubow et al. actually followed from their neglect of radial gradients in the streamline ellipticity.

In this paper, we study the *global* aspects of the *local* elliptical instability in accretion disks using a linear analysis. In §2, we review the linearized equations used and introduce models and assumptions. In §3, we describe the results of the linear analysis focusing on global properties of the instability and the role of radial boundaries. Summary and conclusions follow in §4.

## 2. Equations and Methods

We are concerned with small perturbations of a basic state that is stationary in a frame corotating with the binary and centered on the primary star. The streamlines of the basic flow are closed in this frame, but they are noncircular because of the tidal field exerted by the secondary star. Following G93 and RG, we adopt a generalized form of the shearing sheet approximation (Goldreich & Lynden-Bell 1965; Julian & Toomre 1966). The fluid equations are written in a pseudo-Cartesian coordinate system $xyz$ whose origin orbits with the basic flow along a fiducial streamline. Spatial gradients of the basic flow such as shear and vorticity are retained to the lowest nontrivial order in the size of the patch divided by the mean radius of the fiducial streamline ($r_0$). Strains associated with the ellipticity of the streamlines are included to first order in the tidal field.

It may be asked how we can adopt a local approximation such as the shearing sheet when we are concerned with global questions such as radial boundaries and radial variations in the tidal strain. The answer is that we do not want to model any particular disk in realistic detail, but rather to explore the effects of generic radial inhomogeneities on the growth rates. This we can accomplish by imposing artificial but rapid radial variations in the tidal terms of the shearing sheet equations.

Absent such artificial variations, the fluid equations are (RG)

$$\frac{D\rho}{Dt} + \rho \left( \frac{\partial v_x}{\partial x} + \frac{\partial v_y}{\partial y} + \frac{\partial v_z}{\partial z} \right) = 0, \tag{4}$$

$$\frac{Dv_x}{Dt} + \frac{1}{\rho}\frac{\partial p}{\partial x} - 2\Omega v_y = -2\Omega \left( c_{xx}v_x + c_{xy}v_y \right) + \left( b_{xx}\frac{1}{\rho}\frac{\partial p}{\partial x} + b_{xy}\frac{1}{\rho}\frac{\partial p}{\partial y} \right), \tag{5}$$

$$\frac{Dv_y}{Dt} + \frac{1}{\rho}\frac{\partial p}{\partial y} + 2\left(\Omega + A\right)v_x = -2\Omega \left( c_{yx}v_x + c_{yy}v_y \right) + \left( b_{yx}\frac{1}{\rho}\frac{\partial p}{\partial x} + b_{yy}\frac{1}{\rho}\frac{\partial p}{\partial y} \right), \tag{6}$$

$$\frac{Dv_z}{Dt} + \frac{1}{\rho}\frac{\partial p}{\partial z} - \frac{\partial w_0}{\partial z} = -2\Omega c_{zz}v_z + b_{zz}\left( \frac{1}{\rho}\frac{\partial p}{\partial z} - \frac{\partial w_0}{\partial z} \right), \tag{7}$$

$$\frac{Dp}{Dt} + \gamma p \left( \frac{\partial v_x}{\partial x} + \frac{\partial v_y}{\partial y} + \frac{\partial v_z}{\partial z} \right) = 0, \tag{8}$$

where

$$\frac{D}{Dt} = \frac{\partial}{\partial t} + v_x\frac{\partial}{\partial x} + \left(v_y + 2Ax\right)\frac{\partial}{\partial y} + v_z\frac{\partial}{\partial z}. \tag{9}$$

Here $\Omega$ is the average angular velocity of the basic flow as it travels along the fiducial streamline, and

$$A \equiv \left\langle \frac{r}{2}\frac{d\Omega}{dr} \right\rangle_0 \tag{10}$$

is the average shear.

The coefficients $c_{ij}$ and $b_i$, and the the enthalpy of the basic state,

$$w_0(t) = f_0(t) + f_2(t)z^2, \tag{11}$$

describe the local effects of the tide on the streamlines. In the absence of the tide, the $c$'s and $b$'s would vanish, and $w_0 = (\Omega^2/2)(H^2 - z^2)$. The time $t$ upon which these coefficients depend is really orbital phase along the fiducial streamline. With a predominantly $m = 2$ tidal pattern, the coefficients ($c_{ij}(t)$, $b_i(t)$, $f_0(t)$, and $f_2(t)$) oscillate sinusoidally at frequency $2(\Omega - \Omega_b)$, with an amplitude that depends on ($M_s/M_p$) and ($r_0/r_b$).

Thus $t$ plays a double role in our shearing sheet equations. For the basic state it is just orbital phase. But for perturbations of that state it is a true time variable, since such perturbations need not be periodic. This duality is consistent with our local approximation.

When linearizing equations (4) to (9), we make a couple of simplifications.

First, all derivatives with respect to the azimuthal coordinate ($y$) are dropped, but all three components of velocity are included. In the linear regime, as explained in §§1&4, the $m = 2$ tidal pattern excites primarily $m = 1$ internal waves. The azimuthal wavelength of these waves ($2\pi r_0$) is effectively infinite compared to the size of our local coordinate patch.

Second, the vertical structure of the stationary disk has been neglected by dropping the terms involving $\partial w_0/\partial z$. Local analysis (G93, Vishniac & Zhang (1995)) indicates that these terms are not important in determining the linear growth rate except the radius at which

$$(\Omega - \Omega_b)^2 = (\gamma + 1)\Omega^2, \tag{12}$$

where a vertical pulsation mode resonates directly with the tidal force (Lubow 1981). On the other hand, this simplification makes the mathematical analysis much easier by permitting a Fourier decomposition along the vertical ($z$) direction. Vertical structure and vertical boundary conditions have been considered previously by Lubow, Pringle, & Kerswell (1993) and Vishniac & Zhang (1995). Our focus here is on radial boundary conditions and radial variations in the tidal terms. We therefore assume perturbations of the form

$$\delta q(x, z, t) = \delta q(x, t)e^{ik_z z}. \tag{13}$$

Since we want to study radial inhomogeneities in the basic state, we do not assume a Fourier decomposition along $x$.

Our linearized perturbation equations are

$$\frac{1}{\rho_0}\frac{\partial \delta\rho}{\partial t} + \frac{\partial \delta v_x}{\partial x} + ik_z\delta v_z = 0, \tag{14}$$

$$\frac{\partial \delta v_x}{\partial t} + \frac{c_s^2}{\rho_0}\frac{\partial \delta\rho}{\partial x} - 2\Omega\delta v_y = -2\Omega\left(c_{xx}\delta v_x + c_{xy}\delta v_y\right) + b_{xx}\frac{c_s^2}{\rho_0}\frac{\partial \delta\rho}{\partial x}, \tag{15}$$

$$\frac{\partial \delta v_y}{\partial t} + 2\left(\Omega + A\right)\delta v_x = -2\Omega\left(c_{yx}\delta v_x + c_{yy}\delta v_y\right) + b_{yx}\frac{c_s^2}{\rho_0}\frac{\partial \delta\rho}{\partial x}, \tag{16}$$

$$\frac{\partial \delta v_z}{\partial t} + ik_z c_s^2\frac{\delta\rho}{\rho_0} = -2\Omega c_{zz}\delta v_z + b_{zz}ik_z c_s^2\frac{\delta\rho}{\rho_0}, \tag{17}$$

where the quantities with the subscript $_0$ are those in the unperturbed stationary disk and $c_s$ is the background sound speed. Adiabatic perturbations

$$\frac{\delta p}{p_0} = \gamma \frac{\delta \rho}{\rho_0} \qquad (18)$$

have been assumed, with $\gamma = 5/3$.

We adopt units of length and time such that $\Omega = 1$ and $k_z = 2\pi$. Since we assume a Keplerian disk, $A = -(3/4)\Omega$. The description of the basic state is completed, apart from the tidal distortion, by specifying the sound speed. We have set $c_s = 3$. Note that as we increase $c_s k_z / \Omega \sim k_z H$, inertial waves with frequencies $\lesssim \Omega$ become more nearly incompressible.

The global growth of perturbations depends on the global shape of the tidal distortion. Therefore we allow artificial variations in the tidal coefficients $(c_{ij}, b_i)$ along the radial ($x$) direction. We consider two different cases: (a) the tidal strength is constant in $x$ and (b) the tidal strength has a Gaussian profile,

$$\exp\left[-\left(\frac{x - x_0}{\Delta x_1}\right)^2\right], \qquad (19)$$

with $x_0 = 5$ and $\Delta x_1 = 1.5$; our computational grids extend from $x = 0$ to $x = 10$. The strength and the oscillation period of the tidal distortion scale by an overall factor that is roughly linear in $M_s/M_p$ but has a more complicated dependence on $r_0/r_b$; cf. RG. We fix $r_0/r_b$ at 0.4 and take $M_s/M_p$ in the range $0.05 - 0.2$.

For a given locally homogeneous tide—that is, for $(c_{ij}, b_i)$ independent of $x$—the local growth rate is largest for waves whose natural frequencies lie in exact parametric resonance with the tide. The meridional wavenumbers $(k_x, k_z)$ of such modes satisfy

$$k_z^2 = k_x^2 \frac{(\Omega - \Omega_b)^2}{\Omega^2 - (\Omega - \Omega_b)^2} + \frac{(\Omega - \Omega_b)^2}{c_s^2}, \qquad (20)$$

(see RG). For our choices of $c_s$, $\Omega_b$, and $\Omega(r_0)$, the corresponding ratio of $k_x/k_z$ lies between 0.916 and 0.955.

We have used two different radial boundary conditions, reflecting and absorbing. The reflecting condition is $\partial_x \delta \rho = \partial_x \delta v_y = \partial_x \delta v_z = \delta v_x = 0$. The absorbing condition is created by adding a term. Since it appears that these waves cannot be absorbed at a single grid point or pair of grid points, the absorbing layer is created by adding the term

$$-\frac{\delta \vec{v}}{\tau} \qquad \text{with} \qquad \frac{1}{\tau} = \frac{1}{\tau_0}\left(1 - \frac{x}{\Delta x_2}\right) \qquad \text{for} \qquad 0 \le x \le \Delta x_2, \qquad (21)$$

to the right hand side of equations (15)-(17). We have used $\tau_0 = 4$ and $\Delta x_2 = 2$. The round-trip attenuation of a wave packet traveling at group velocity $v_g$ from $x = \Delta x_2$ to $x = 0$ and back again is

$$\Delta \ln \delta v \approx -2 \int_0^{\Delta x_2} \frac{dx}{v_g \tau(x)} \approx -5.9, \qquad (22)$$

assuming $v_g = 0.085$ (see below). Note also that $k_x \Delta x_2 \approx 12$ for the fastest-growing modes in our simulations, so the damping can be estimated with a WKB formula like (22). Thus, the waves of interest are almost completely absorbed in the layer $0 \leq x \leq \Delta x_2$.

The linearized perturbation equations (14)-(17) constitute an initial-value problem. We have used a localized initial perturbation,

$$
\begin{aligned}
\delta v_z &= \begin{cases} A \sin\left[2\pi(x - 4.5)\right], & \text{in } 4.5 \leq x \leq 5.5 \\ 0, & \text{otherwise,} \end{cases} \\
\delta p &= \delta\rho = \delta v_x = \delta v_y = 0,
\end{aligned}
\tag{23}
$$

with $A = 10^{-20}$. We have evolved the equations using the Runge-Kutta-Verner fifth and sixth order method with 1000 equally spaced cells along $0 \leq x \leq 10$.

## 3. Results

In the calculations with a spatially constant tidal distortion, the perturbations grow exponentially with localized disturbances slowly spreading out with group velocity (G93)

$$
v_g \approx -\text{sign}(k_r) \frac{\Omega}{|k_z|} \left(1 - \frac{\Omega_b}{\Omega}\right) \left[1 - \left(1 - \frac{\Omega_b}{\Omega}\right)^2\right]^{1/2},
\tag{24}
$$

which is $\approx 0.085$ for our parameters. Fig. 1 shows the temporal evolution of the perturbation quantities averaged in quadrature over the grid in a calculation with reflecting boundaries on both sides. Here, the strength and the oscillation period of the tidal are appropriate for $M_s/M_p = 0.2$ and $r_0/r_b = 0.4$, thus $\Omega_b/\Omega \approx 0.27$. From this plot, the measured growth rate is $s = 0.139$, in good agreement with the local growth rate predicted by G93. Fig. 2 shows the velocity eigenfunctions at four epochs. Evidently the perturbations spread out not at the sound velocity ($c_s = 3$) but at the much smaller inertial-wave group velocity (24). It is interesting to note that as the perturbations grow, the dominant $x$ and $z$ wavenumbers come to have the approximate ratio $k_x/k_z \approx 0.93$, close to the one which gives the maximum local growth equation (20).

When we repeat these calculations with an absorbing layer rather than a reflecting boundary on the left side, we see a nearly identical exponential growth rate. In Fig. 3, we have plotted the velocity eigenfunctions for a calculation with the same tidal parameters as in Figs. 1&2. The wave is clearly suppressed in the absorbing layer, but otherwise it looks very similar to that of Fig. 2. We have also done the calculation with absorbing layers at both boundaries (not shown); the effect on the growth rate is negligible. Evidently, the instability does not require reflecting radial boundaries.

Next we take a tidal distortion that varies along the radial direction according to the gaussian profile (19). The strength of the tidal terms at the boundaries ($x = 0$ and 10) is $\exp[-(5/1.5)^2] = 1.5 \times 10^{-5}$ times smaller than the peak strength at $x = 5$, and the local growth rate is reduced in direct proportion. We take a reflecting boundary on the right and an absorbing layer on the left.

From Fig. 4, the global growth rate is $s = 0.116$, which is comparable to, but significantly smaller than, the rate observed previously.

Fig. 5 shows the velocity perturbations at four different epochs. Since the wave has the same shape in the last three epochs, we may consider it a global mode. The eigenfunction is strongly peaked at the center of the grid where the local growth rate is largest. However the wings of the eigenfunction are clearly growing at the same exponential rate as the central parts, not at their own local growth rate. We can fit the eigenfunction to an envelope of the form

$$|\delta v_x|_{\text{max}}, \ |\delta v_y|_{\text{max}}, \ |\delta v_z|_{\text{max}} \propto \exp\left(-\left|\frac{x-5}{v_{\text{g}}}\right|\right), \tag{25}$$

with $v_{\text{g}} \approx 0.8$. Notice that this envelope differs in shape from the gaussian profile of the local growth rate (19).

Our interpretation of the results in Figs. 4&5 is that the wave grows locally in the central region where the local growth rate is largest; however, because the mode is also a propagating wave, it leaks out into the surrounding regions where the local growth rate is negligible. The amplitude at time $t$ in the wings $\delta x$ from the center is comparable to the central amplitude at the retarded time $t - |\delta x / v_{\text{g}}|$; this explains the envelope shape (25).

On this interpretation, the global growth rate should depend on the maximum local growth rate, the size of the rapid growth region, and the local group velocity. If disturbances escape from the growth region in less than a nominal e-folding time, we expect the instability to be suppressed. In order to test and quantify this hypothesis, we have done several calculations with the same gaussian profile (19) for the tidal terms, but with different maximum strengths. The latter were chosen as appropriate for $r_0/r_{\text{b}} = 0.4$ and $M_{\text{s}}/M_{\text{p}} = 0.16, 0.13, 0.1, 0.08$, and $0.05$, as well as $0.2$. In Fig. 6, we have plotted the global growth rate $s$, normalized to the maximum local growth rate $s_{\text{max}}(r_0, M_{\text{s}}, M_{\text{p}})$, against $(v_{\text{g}}/Ls_{\text{max}})$. Here $L$ is a measure of the size of the rapid growth region,

$$L = \int \frac{s_{\text{loc}}}{s_{\text{max}}} dx = \int_{-\infty}^{\infty} \exp\left[-\left(\frac{x-x_0}{\Delta x_1}\right)^2\right] dx \approx 2.66. \tag{26}$$

The global growth rate is well fit by

$$\frac{s}{s_{\text{max}}} \approx 1 - \frac{v_{\text{g}}}{Ls_{\text{max}}}. \tag{27}$$

Thus the global growth rate is comparable to the maximum local rate unless $v_{\text{g}} \gtrsim Ls_{\text{max}}$, in which case instability is quenched. With appropriate values for $L$ and $s_{\text{max}}$, this formula may be used to estimate the global growth rate in accretion disks. We suspect that in cases where the local rate peaks in an extremely narrow annulus, the value of $s_{\text{max}}$ to be used in this formula should be obtained by averaging over a radial range $\Delta r \sim 1/k_x$ around the peak.

## 4. Summary and Conclusions

We have shown that the excitation of perturbations by the elliptical instability in tidally distorted disks is essentially local, and is insensitive to boundary conditions as long as the unstable region is far enough from the boundaries. If the local growth rate varies radially then regions of rapid growth launch waves into the regions of slower growth. The global growth rate is proportional to the maximum local growth rate, but is reduced by the ratio of the group velocity to the size of the rapid-growth region, i.e. by the rate at which disturbances escape the region. We conclude that the tidal excitation of internal waves is insensitive to radial boundary conditions. Since previous work has shown that it is only moderately sensitive to vertical structure (Vishniac & Zhang (1995)) it follows that this process will inevitably occur in tidally distorted accretion disks. Given the sensitive dependence of the growth rate on radius, the torque exerted by the creation of the internal waves is likely to play a major role in determining the outer boundary of a disk.

Our results also imply that it is possible for the $m = 2$ component of the tidal stresses to excite waves with $m \neq 1$, although not as easily as it will drive waves with $m = 1$. A wave with some value of $m$, not necessarily equal to 1, will drive a second wave with $k_{r2} = -k_{r1}$, $\bar{\omega}_2 = 2(\Omega - \Omega_{\rm b}) - \bar{\omega}_1$, $m_2 = 2 - m_1$, and $k_{z2} = -k_{z1}$. The second wave can, in turn, beat against the tidal stresses to reinforce the first wave. As long as $m_1/r$ and $m_2/r$ are small compared to $k_{z1}$ and $k_{r1}$ these resonant conditions imply that $\bar{\omega}_1 = \bar{\omega}_2$. Since $\bar{\omega}$ is a different function of $r$ for waves with different values of $m$ the resonant conditions can only be satisfied over some annulus of size $\Delta r$ such that

$$\left| \int_{r_0}^{r_0 + \Delta r} (k_{r1} + k_{r2}) dr \right| \leq 1. \tag{28}$$

If we assume that $k_{r1} + k_{r2} = 0$ in the middle of this annulus then

$$k_{r1} + k_{r2} = \frac{\partial k_r}{\partial \bar{\omega}} \Delta \bar{\omega} = v_{\rm g}^{-1} \left( -\frac{3}{2} \Omega \frac{(m_1 - m_2)}{r} \right) (r - r_0 - \frac{\Delta r}{2}). \tag{29}$$

Equations (28) and (29) imply an upper limit on $\Delta r$ of

$$\Delta r \leq \left( \frac{8 r v_{\rm g}}{3 \Delta m \Omega} \right)^{1/2}, \tag{30}$$

where $\Delta m \equiv |m_1 - m_2|$. Following equation (27) we see that the condition for the existence of an elliptical instability involving a wave pair $(m_1, 2 - m_1)$ is

$$\frac{3 \Delta m \Omega v_{\rm g}}{8 r} \leq s_{\rm max}^2. \tag{31}$$

This condition can always be satisfied for $v_{\rm g}$ very small, i.e. for very large wavenumbers; in fact, using equation (2) for $s_{\rm max}$ and (24) for $v_{\rm g}$, we find when $r \ll r_{\rm b}$ that

$$\Delta m \lesssim 2 k_z H \left( \frac{M_{\rm s}}{M_{\rm p}} \right)^2 \left( \frac{r}{10 H} \right) \left( \frac{r}{0.4 r_{\rm b}} \right)^{21/4}. \tag{32}$$

On the other hand, such waves will have very small saturation amplitudes. If we restrict ourselves to large scale modes, i.e. $k_z H$ of order unity, then equation (31) is satisfied for $\Delta m \neq 0$ only in the outer parts of thin disks.

This mechanism drives standing waves, i.e. waves with equal amplitude in outward and inward propagating waves. Nevertheless, since the growth rate varies with radius as $r^{3/2}$ far from the vertical resonance, and much more steeply near it, the local wave field will in general be asymmetric, with more waves moving down the local gradient in $s$ than moving up it. This implies that at small radii we expect to be dominated by inward propagating internal waves with $m = 1$, carrying negative angular momentum. This will produce a positive angular momentum flux whose amplitude will depend on whether the amplitude of the waves is limited by nonlinear interactions between these waves or by interactions with some independently generated local turbulence. Vishniac & Diamond 1989 assumed the former and estimated an effective (nonlocal) dimensionless viscosity $\alpha \sim (H/r)^2$. Whether or not this effect is significant depends on the nature of competing processes.

This work was supported in part by the Non-Directed Research Fund of Korea Research Foundation 1993 (DR), and by NASA grants NAGW-2419 and NAG5-2796 (JG) and NAG5-2773 (ETV).

## FIGURE CAPTIONS

**Fig. 1** Evolution of spatially averaged r.m.s. velocity perturbations, assuming a spatially constant tidal distortion appropriate for $M_s/M_p = 0.2$, $r_0/r_b = 0.4$. Reflecting radial boundary conditions.

**Fig. 2** Velocity waveform at four epochs for the same calculation as in Fig. 1.

**Fig. 3** Like Fig. 2, but with a reflecting boundary at the right and and an absorbing layer on the left [cf. Eqn. (21)]. Note that amplitude is almost identical with Fig. 2 at corresponding times except in the absorbing layer $0 \leq x \leq 2$.

**Fig. 4** Evolution of spatially averaged perturbations for tidal distortion with peak strength as in Figs. 1-3 but modulated by the gaussian profile (19). Reflecting and absorbing boundaries as in Fig. 3.

**Fig. 5** Waveforms at four epochs for the same calculation as in Fig. 4. Note that abscissa is logarithmic rather than linear in the final panel; compare equation (25).

**Fig. 6** The global growth rate normalized with the maximum local growth rate ($s/s_{\max}$) as a function of the ratio of the group velocity to the size of the rapid growth region times the maximum local growth rate ($v_g/Ls_{\max}$). Tidal terms modulated by gaussian (19), with peak strengths and oscillation periods for $M_s/M_p = 0.2$, 0.16, 0.13, 0.1, 0.08, and 0.05 and $r_0/r_b = 0.4$. Boundary conditions as in Figs. 3-5.

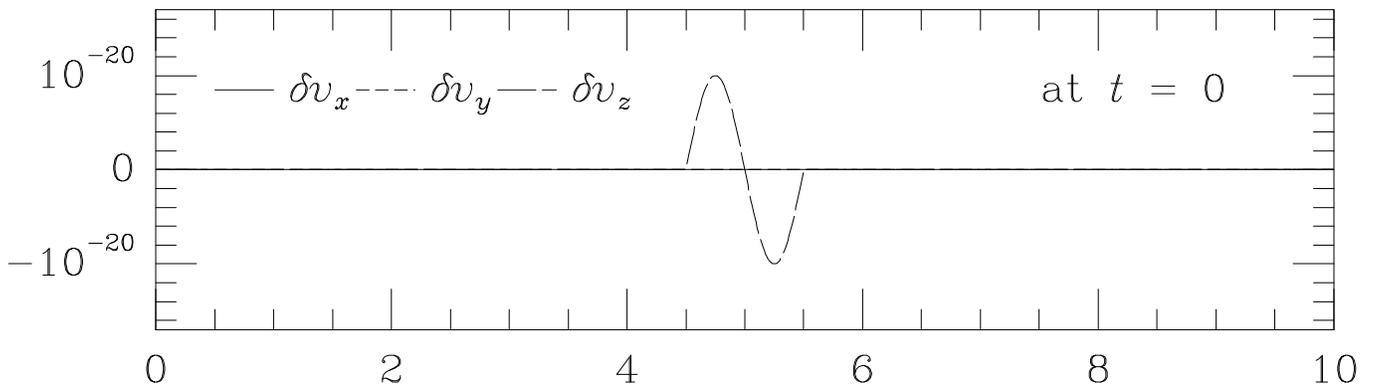
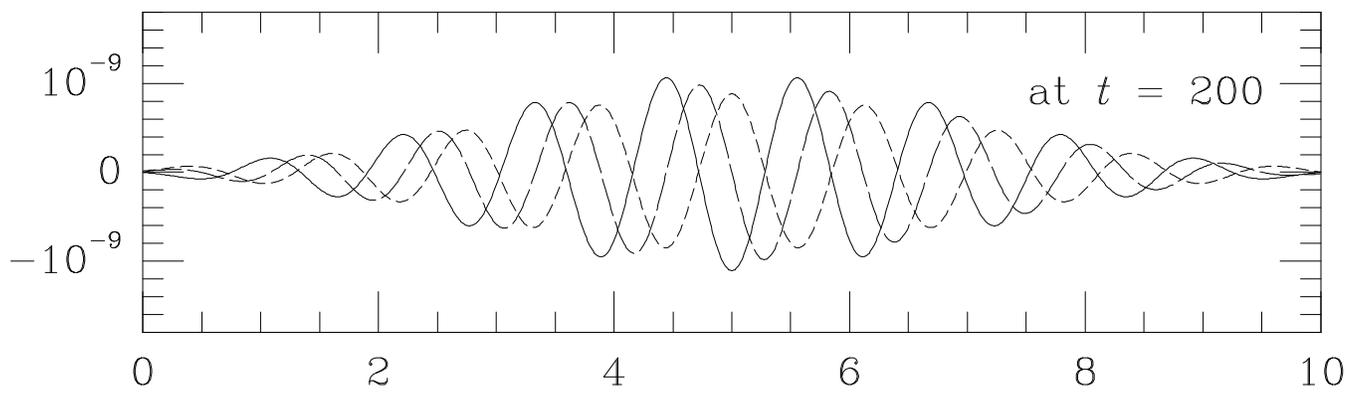
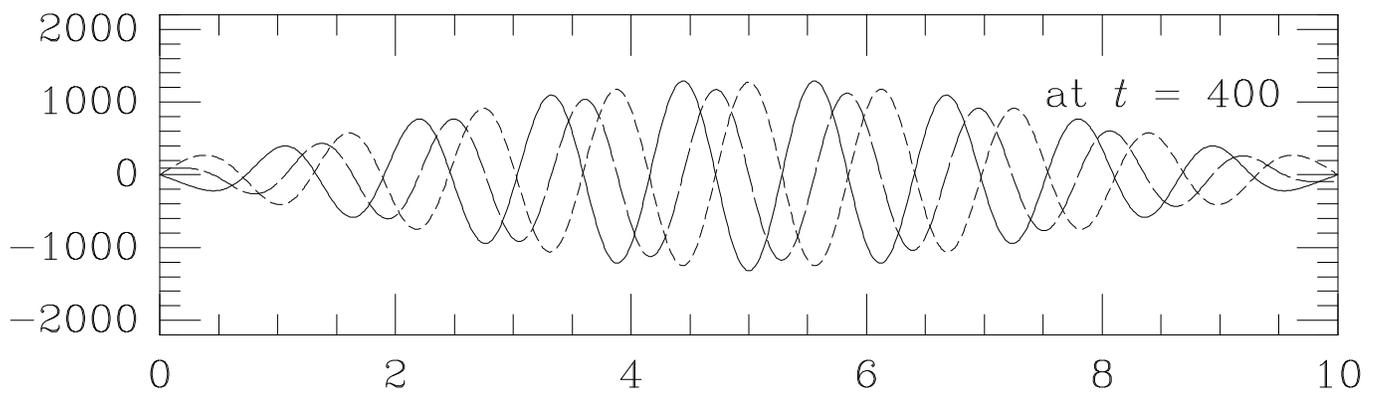
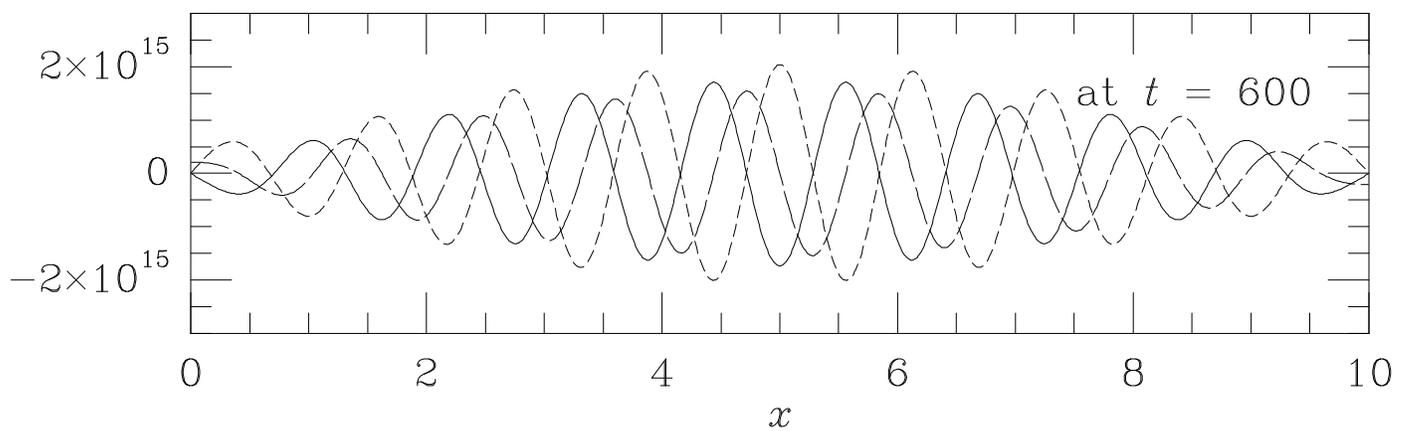

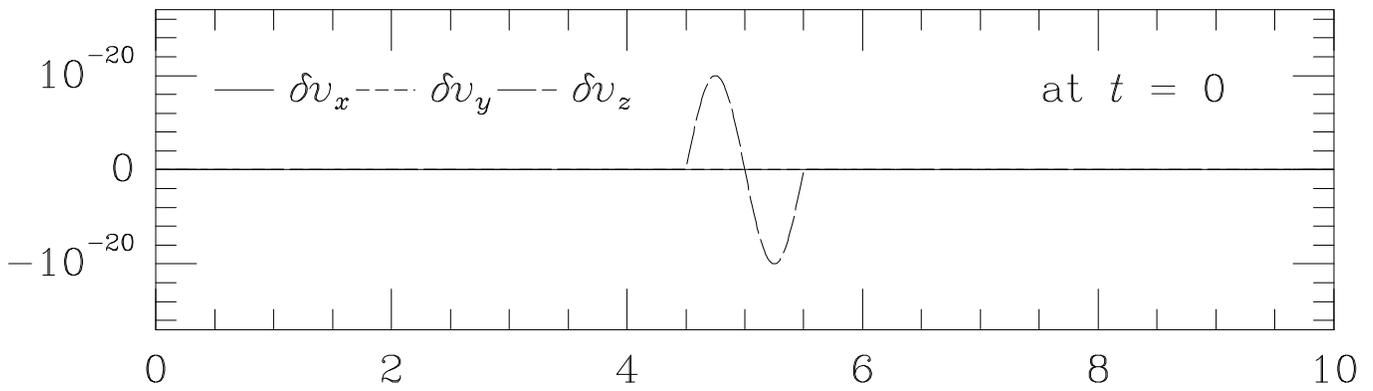

at $t = 0$

$\delta v_x$ —— $\delta v_y$ ----  $\delta v_z$ – –

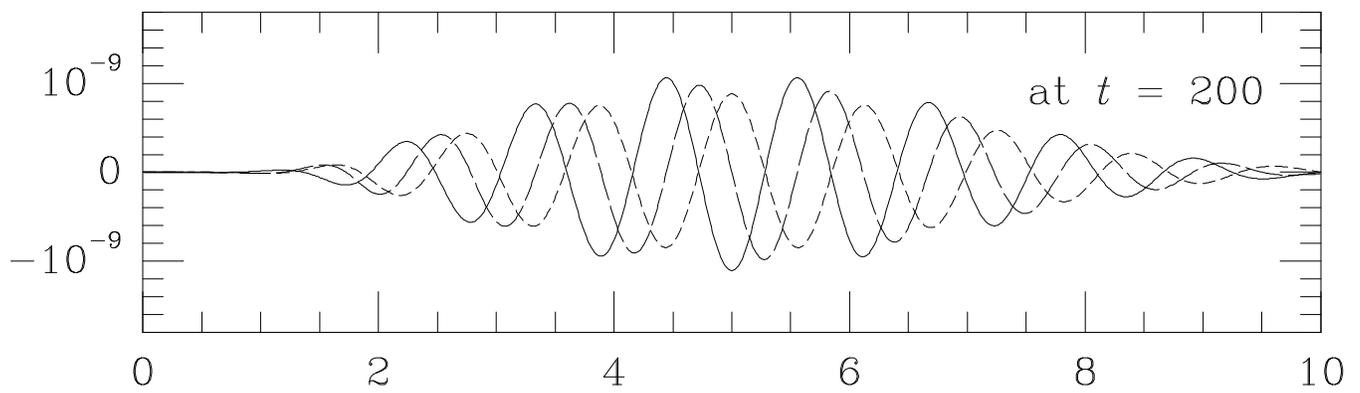

at $t = 200$

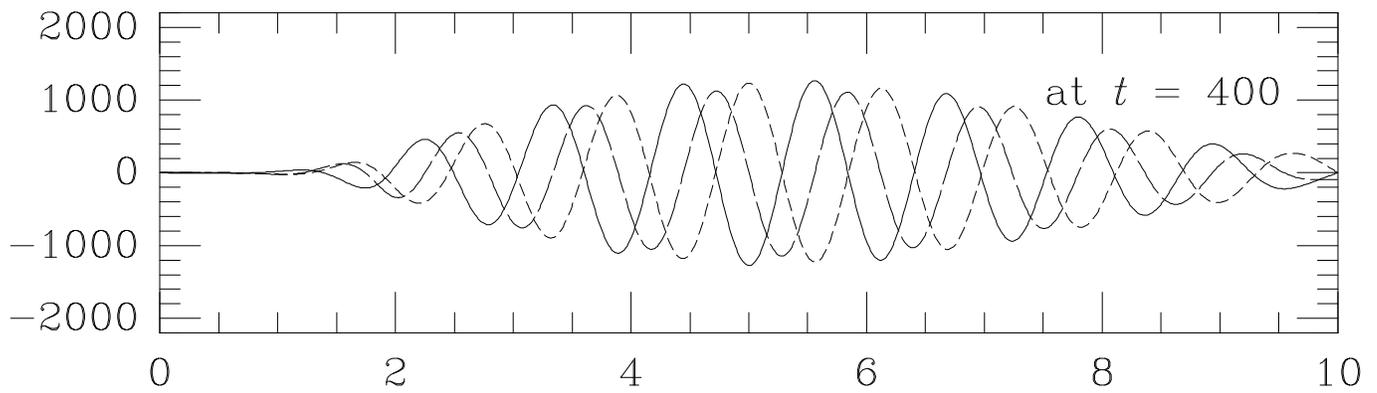

at $t = 400$

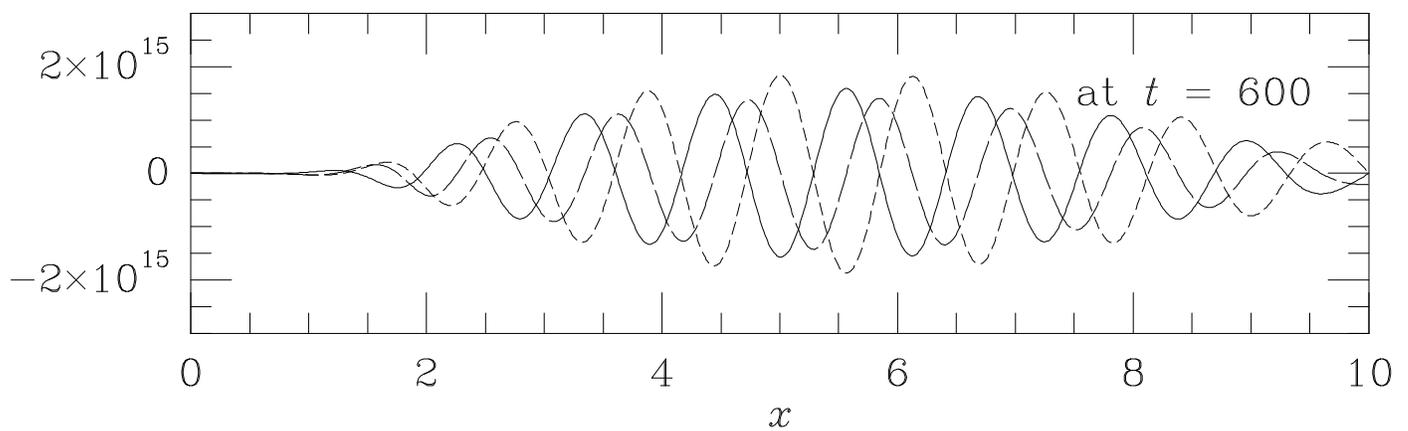

at $t = 600$

$x$

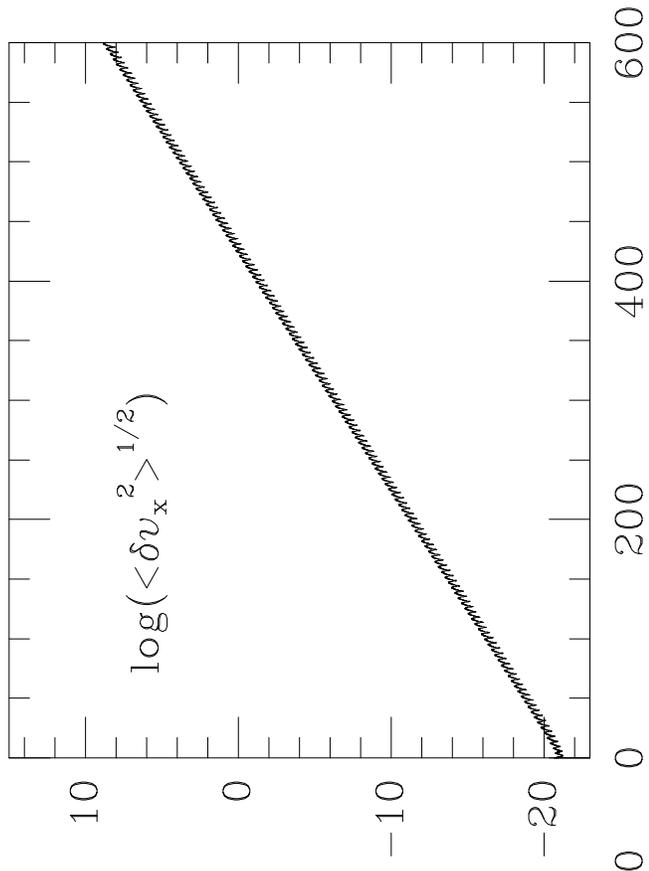

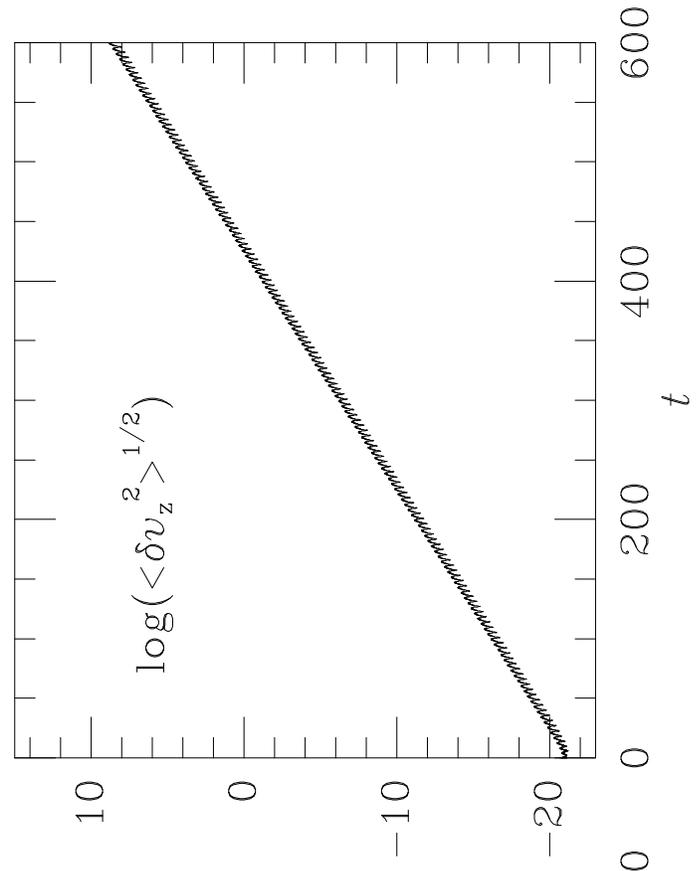

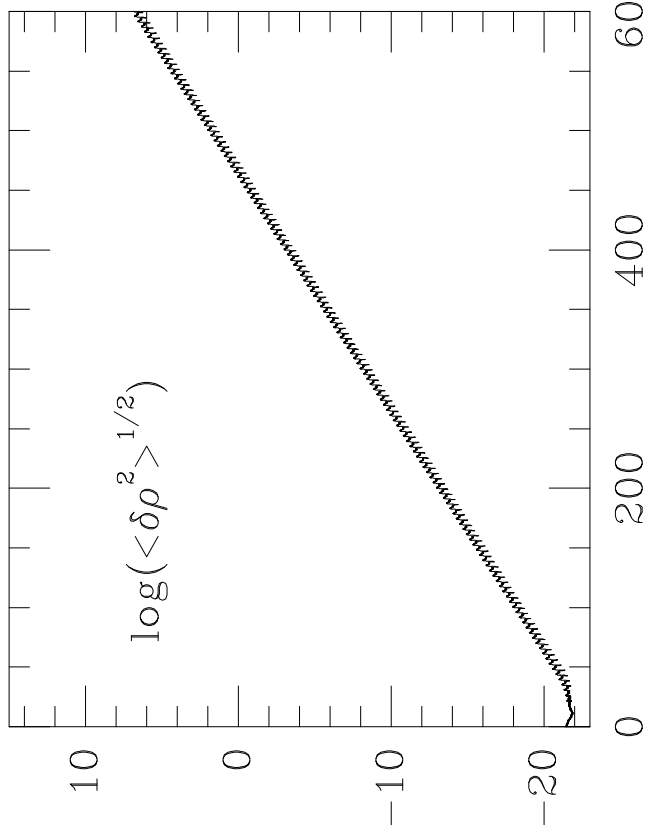

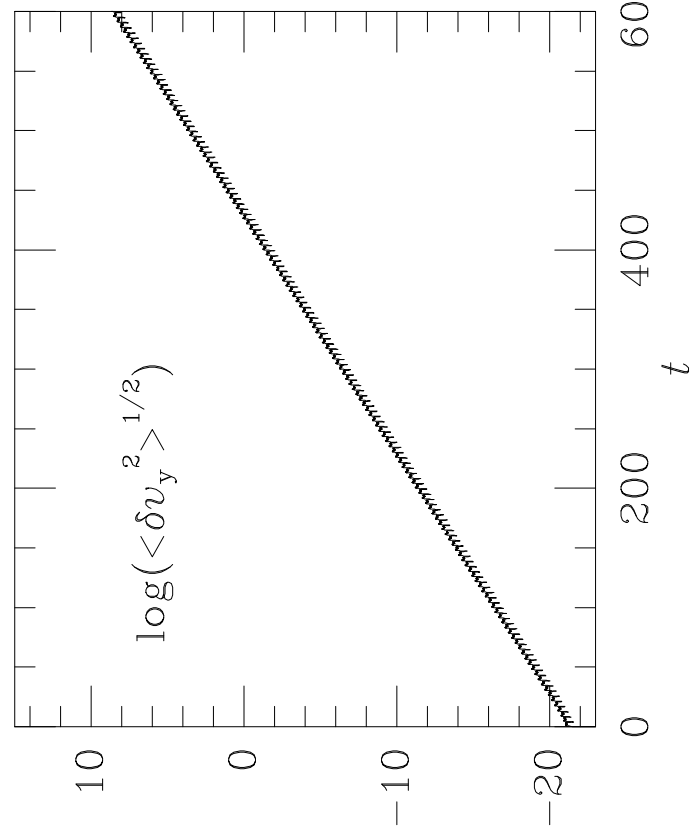

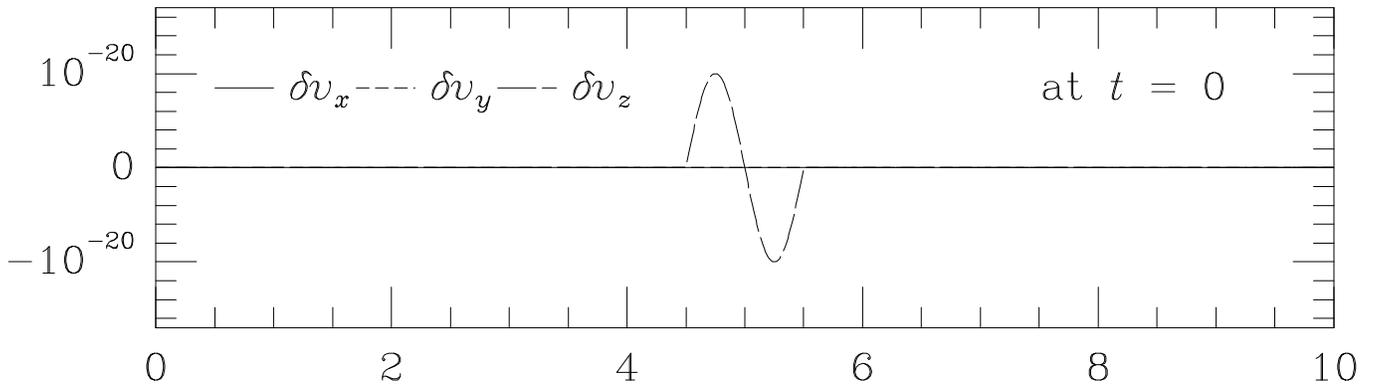

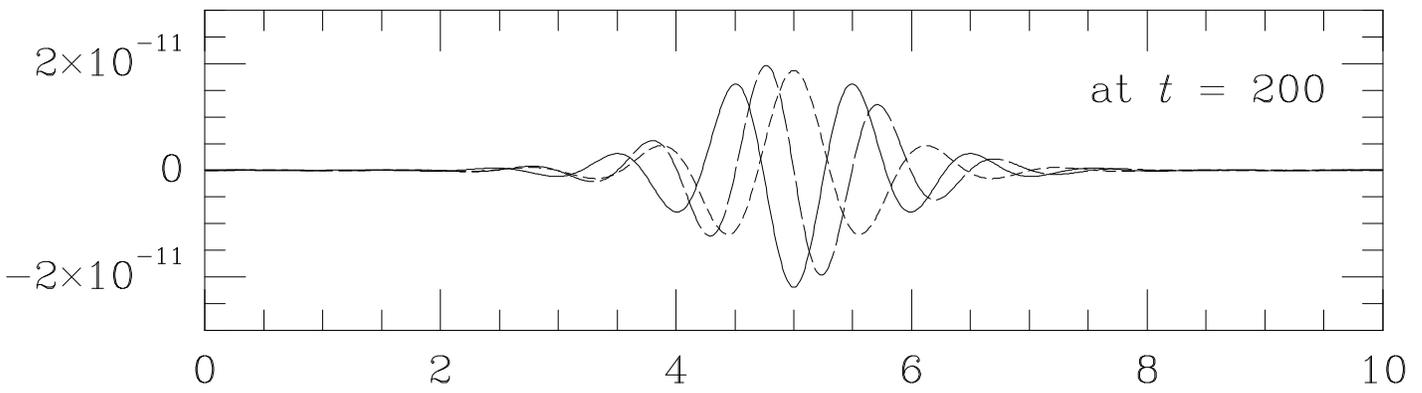

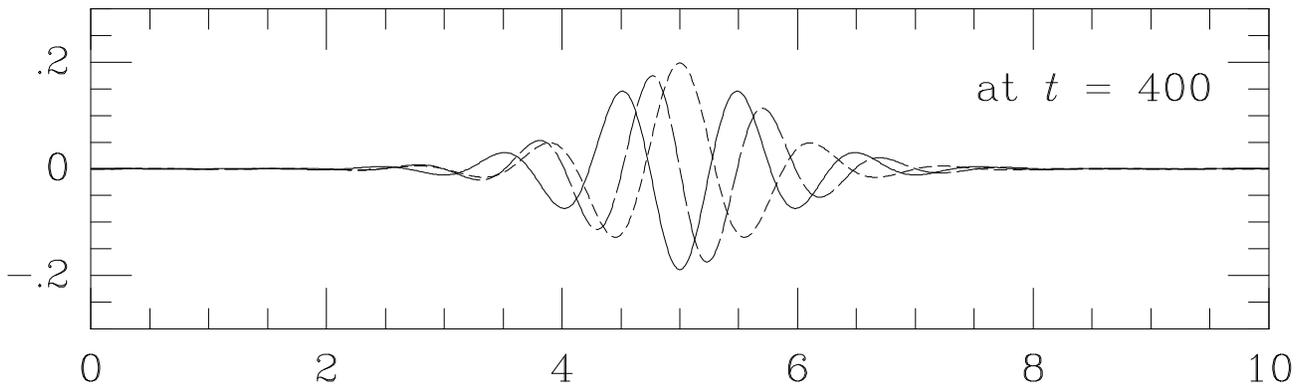

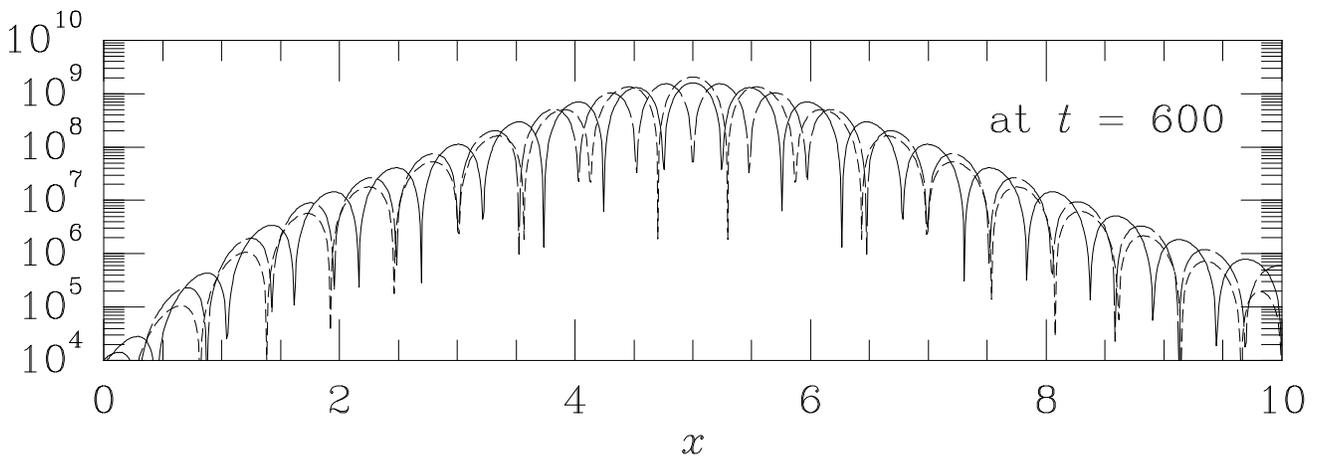

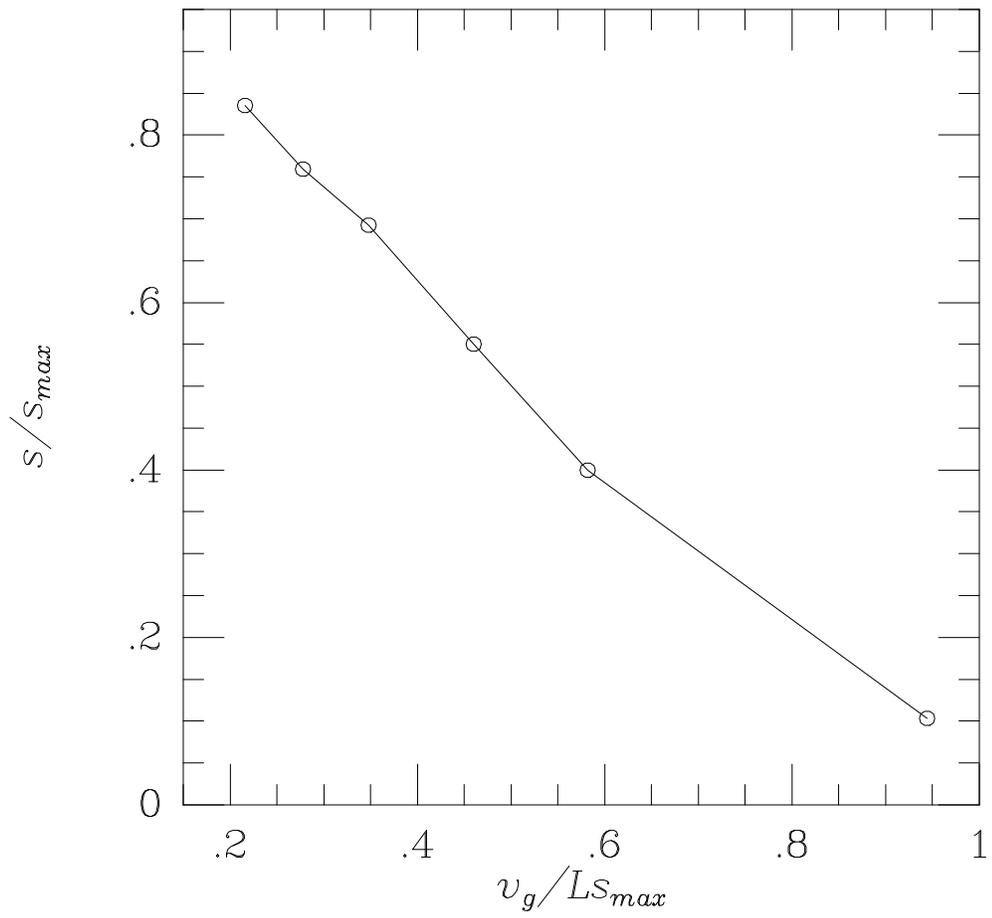